\documentclass[runningheads]{llncs}

\usepackage{framed}
\usepackage{todonotes}
\usepackage{subfig}
\usepackage{url}
\usepackage{graphicx}
\usepackage{lscape}
\begin{document}
\title{Building the MSR Tool Kaiaulu: Design Principles and Experiences}
%
%
\author{Carlos Paradis\inst{1,2}\orcidID{0000-0002-3062-7547} \and
Rick Kazman\inst{1,2}\orcidID{0000-0003-0392-2783}}
\authorrunning{C. Paradis et al.}
%
\institute{University of Hawaii at Manoa, Honolulu HI 96822, USA \and
\email{\{cvas,kazman\}@hawaii.edu}}
%
\maketitle              
\begin{abstract}
Background: Since Alitheia Core was proposed and subsequently retired, tools that support empirical studies of software projects continue to be proposed, such as Codeface, Codeface4Smells, GrimoireLab and SmartSHARK, but they all make different design choices and provide overlapping functionality.
Aims: We seek to understand the design decisions adopted by these tools--the good and the bad--along with their consequences, to understand why their authors reinvented functionality already present in other tools, and to help inform the design of future tools.
Method: We used action research to evaluate the tools, and to determine a set of principles and anti-patterns to motivate a new tool design.
Results: We identified 7 major design choices among the tools: 1) Abstraction Debt, 2) the use of Project Configuration Files, 3) the choice of Batch or Interactive Mode, 4) Minimal Paths to Data, 5) Familiar Software Abstractions, 6) Licensing and 7) the Perils of Code Reuse. Building on the observed good and bad design decisions, we created our own tool architecture and implemented it as an R package.
Conclusions: Tools should not require onerous setup for users to obtain data. Authors should consider the conventions and abstractions used by their chosen language and build upon these instead of redefining them. Tools should encourage best practices in experiment reproducibility by leveraging self-contained and readable schemas that are used for tool automation, and reuse must be done with care to avoid depending on dead code.
\end{abstract}
\section{Introduction}

Research into quality dimensions of software project  requires the analysis of large quantities of data. For researchers this typically means mining data from multiple open source software projects.
Pre-processing data, calculating metrics and flaws, and synthesizing
composite results from a large corpus of project artefacts
is a tedious and error prone task lacking immediate scientific
value \cite{Gousios:2009}---it is seen merely as a means to an end. This was the motivation for the Alitheia Core \cite{Gousios:2009}, which was made available in 2009 for the software engineering community. It provided  features for data collection, integration and analysis services and emphasized an easy to use extension mechanism. Yet, as of today, Alitheia Core is a dormant (read-only) project in   GitHub\footnote{\url{https://github.com/istlab/Alitheia-Core}} and several other tools replicate at least some of its functionality. 

What went wrong? Why have many tools re-implemented the same ``tedious and error prone'' tasks the Alitheia Core? And do the current tools live up to the promise of Alitheia Core? In this work, we revisit lessons learned by the Alitheia Core authors and the design choices made by the other more recent tools using an action research \cite{Easterbrook:2008} approach. 

Our contributions in this paper are twofold: first, we present a set of key design decisions derived from an analysis of the aforementioned tools which either facilitated or hindered reusability, reproducibility, interoperability and extension of functionality. Second, we present our tool, Kaiaulu\footnote{The documentation for the tool can be found at \url{https://github.com/sailuh/kaiaulu}}, which builds upon the design decisions made from these prior tools, and which we believe fills a gap in the existing mining software repositories ecosystem.

\section{Studied Tools and Lessons Learned}

The tools that we studied are Codeface \cite{Joblin:2017}, Codeface4Smells \cite{Tamburri:2019}, GrimoireLab \cite{Moreno:2019,Duenas:2018} SmartSHARK, \cite{Trautsch:2018,Trautsch:2020} and PyDriller \cite{Spadini:2018}. We now present our observations regarding the strengths and weaknesses of these tools in terms of their design choices and note, throughout the work, lessons learned by the authors of Alitheia Core \cite{Gousios:2009}
presented in \cite{Menzies:2016}. Many of these lessons are applicable and worthy of consideration in new tools with similar intents. We employed an action research methodology in studying these tools, but do not describe the details of that research here, due to space limitations.

\subsection{Abstraction Debt}\label{sec:abstraction_debt}

We have observed different levels of abstraction employed in the surveyed tools, ranging from applications that are built as monoliths to those built from smaller components. 
This is consistent with what has been noted in machine learning systems as abstraction debt \cite{Holt:2015}, i.e. a lack of key abstractions to support the functions and growth of MSR tools. 

Codeface was created as a monolithic application, in which an entire project's Git log or mailing list is analyzed.
It abstracts a complete end-to-end pipeline, implemented by a command line interface (CLI), and outputs a database dump of a project. It is therefore difficult for other applications to build on some of its unique features, for example, using its Git log parser that parses at function (rather than file) granularity. 

Both GrimoireLab and SmartSHARK define several components, each with its own CLI, but the component abstractions they employ are not the same. To provide a point of comparison, Grimoire's Lab Perceval provides a CLI to obtain data from many data sources (e.g. GitHub, Git, Bugzilla, Jira, mailing lists, etc), serving as a single interface for data collection. In contrast, SmartSHARK defines its abstraction per data source type and, in the case of data acquisition, at a more fine-grained level than Perceval. For example, consider issueShark and vcsShark, two components of SmartShark. IssueShark defines abstractions for different types of issues tracker sources, and vcsShark for different types of version control systems. SmartSHARK's abstractions facilitate defining additional features specific to a data source type, such as separating static vs. dynamic data in issue trackers (e.g. creation time of the issue vs. comments), regardless of its underlying implementation (e.g. Jira or Bugzilla)\footnote{\url{https://github.com/smartshark/issueSHARK\#introduction}}. 

Pydriller is a single component and is smaller in scope as it only abstracts Git repositories. However it is different from the other tools in that it provides an API instead of a CLI.  Its motivation is also different: it wraps around PythonGit, which in itself provides a Pythonic API to nearly all features of Git, to provide an API catered towards mining software repositories only. In providing just a subset of Git functionality, it exposes functionality catering specifically to the needs of mining repositories.

The decision between choosing a CLI or API has tradeoffs. An issue with command line only interfaces occurs when an end-user may be interested in a different abstraction of the data not preconceived by the authors. However an API requires the user to be familiar with the programming language the tool was built on top of, whereas a CLI does not. 

\textit{From the above we derive the following lessons learned: End-to-end pipelines such as Codeface's  limit the ability of other researchers to build on top of them. Defining more specific abstractions per data type, whether via CLI or API as issueShark and PyDriller do, facilitates building additional functionality specific to a particular data type, or audience. Moreover, CLIs can be built on top of a well-defined API, providing the benefit of both interfaces, as we do in Kaiaulu}.

\subsection{Tool Configuration Files vs Project Configuration Files}\label{sec:config_files}

In \cite{Trautsch:2020}, the authors of SmartSHARK noted that one of their goals was to support replication through the storage of data in a single harmonized schema. Replication, it is argued, is supported by a common dataset. However, we have observed that replication is also being done within configuration files in Codeface.  

Codeface uses a concept we named project configuration files. These files provide a  single compact source where parameters associated with the acquisition and manipulation of a dataset can be stored. Project configuration file parameters are required for tool execution, and they are a pragmatic, lightweight and human-readable way to specify reproducible results. Project configuration files also save time when a project is re-analyzed in other studies, as some project-specific information may not be obvious from the dataset alone.

Of all the tools we have reviewed, only Codeface provides users with a means to specify project configuration files. This led to a large collection of project configurations that have been versioned in Codeface over time\footnote{See \url{https://github.com/siemens/codeface/tree/master/conf} and \url{https://github.com/maelstromdat/codeface4smells_TR/tree/master/Configurations} for Codeface and Codeface4Smells respectively}. This information, which supports repeatability, may otherwise not have been possible (or at least easy) to reconstruct if all that was shared was the data.  

We note that externalizing parameter choices in data acquisition and manipulation tasks has been more prominent in machine learning frameworks, for example to define experiments in configuration files\footnote{\url{https://xnmt.readthedocs.io/en/latest/experiment_config_files.html}}, which include machine learning model selection and choice of model hyper-parameters \cite{Neubig:2018}. 

\textit{From the above, we derive the following lessons: integrating configuration files that are human-readable and leveraged by the tool can enable reproducibility, without the hurdles of sharing large quantities of primary data}.

\subsection{Batch Mode, Interactive Mode, and Literate Programming}\label{sec:batch_interactice_literate}

As we noted before, with the exception of PyDriller, every tool defines a CLI, but not an API. This means the only way to interact with these tools is batch mode. Meanwhile, PyDriller does not offer a CLI, only an API, which confers its users the ability to leverage Python's interactive mode to \textit{explore} the data. However, it does not include a CLI for batch mode processing, for out-of-the-box data acquisition, processing or data analysis. What we observe then is that existing tools decide on either CLI or API, but not both. We believe, however, that the mining of software repositories requires a tool capable of both, supporting an iterative process of data exploration, and when concluded, a way to enact batch processing to scale up. 

To illustrate our claim---as no existing tool provides both capabilities---we provide a few examples: in a recent socio-technical study, we needed to do identity matching, applying heuristics that have been published by other authors (e.g. \cite{Bird:2006,Zhu:2019}) to assign identities to developers who use different names and e-mails in version control systems and mailing lists. Consider the case where we chose the simplest method, where developers whose name or e-mail match are assigned the same id. At first glance, this seems like a reasonable assumption. However, it was due to experimenting interactively with the identity matching API that we discovered that all core developers, due to the use of an issue tracking system, ended up sharing the same e-mail address. We noted this case as a unit test until a better heuristic could be found, and then examined the data for other cases until we were satisfied with the results. We then saved the observed parameters in a project configuration file, and used it to deploy a batch process to collect various computationally intensive architectural metrics. 

We have had similar experience in determining and testing heuristics to filter files in a repository, or determining the method that developers adopt to annotate issue numbers in commit messages. Because each project may apply its own conventions, tools that offer an experimentation capability, and then defer mass data processing to batch more efficiently support the full workflow of a researcher in mining software repositories. 

The described interactive data explorations could certainly have been done in a Python or R session, but it is better to leverage literate programming using, for example, Python or R Notebooks, so that the rationale of the design experiment is not lost. However, care must be taken to not extensively rely on notebooks without further refactoring functionality into the code base, leading to dead experimental code paths \cite{Holt:2015}.

\textit{Our learned lessons here were: existing tools choose either APIs or CLIs (supporting batch or interactive modes). However, making \textit{both} interfaces available will better support users in their various research efforts in mining software repositories. The use of Notebooks to illustrate and explain the API complements the API, provided functionality is not entirely written in Notebooks. In Kaiaulu, we leverage both APIs and Notebooks, which is a common practice in R packages, therefore avoiding abstraction debt}.

\subsection{Minimal Paths to Data}\label{sec:minimal_paths}

According to \cite[p.233]{Menzies:2016}, the effort required to learn how infrastructure code works has to be proportional to the gains and account for deprecation. We agree with this observation. Let us look at how existing tools manage this concern.

When using GrimoireLab components (in particular Perceval) the minimal path to data is surprisingly short. Provided with a Git repository URL, or a local copy, it will output a JSON file to stdout. Likewise, provided with a URL to a website mbox or local file, it will also provide a JSON file to stdout. A developer can easily integrate wrappers to its CLI, and users can easily obtain data for a project of interest. In this ecosystem, a database is available, but it is optional: users need not to concern themselves with learning GrimoireLab's Elastic Search database to obtain data. 

This is in contrast to Codeface and SmartSHARK, both of which require user familiarity with MySQL and MongoDB respectively, along with their data model schemas to obtain the equivalent version control system and mailing list data. The minimal path to data in these cases is much longer, including the setup overhead and integration with other tools. 

When data integration is sought in the database, GrimoireLab retains its approach of keeping the data closest to source, and not harmonizing it in a schema that facilitates integration \cite{Trautsch:2020}. Codeface's MySQL and SmartSHARK's MongoDB provide a harmonized schema, which makes it easier for users to store the various types of data.

In the case of PyDriller, which provides an API, the minimal path to data requires familiarity with the Python programming language. This offers the convenience of reshaping the data to the user's final need, but adds an overhead to the user for familiarization with the API, instead of just the raw data schema from the source of interest (which the user is likely already familiar with for their research purposes).
One researcher \cite[p.39]{Giarola:2016} who extended Codeface4Smells identified a problem of Pipeline Jungles \cite{Holt:2015}, due to heavy reliance on a folder hierarchy and file name conventions.

\textit{Our lessons learned here were: databases need not be a requirement to provide users with various data sources. This also simplifies component reuse by other tools and decreases the likelihood of reinventing the wheel. Providing a minimal path does not exclude providing a database for researchers, as Perceval shows. However providing a harmonized schema  can save researchers from having to re-implement code to integrate the same kinds of infrastructure over and over. Lastly, providing an API gives some flexibility to users to reshape the data with the tool. But user familiarity with the programming language and API is a kind of overhead and this does not seem ideal, as the data could be provided directly via a CLI leaving a task for the researcher to adapt it in their own programming language. As such, we believe having available a CLI that outputs the data as Perceval does, and a harmonized schema as in Codeface and SmartSHARK, provides the best combination}.

\subsection{Other Design Decisions}\label{sec:other_decisions}

We briefly mention here other (more minor) design decisions that we believe may cause difficulties in adoption.

\textbf{Familiar Software Abstractions.}\label{sec:software_abstractions} Both Perceval and PyDriller leverage a common interface for end-users. They are both Python libraries, and provide the expected interactions for CLI and API respectively. In Perceval's CLI, provided with a list of parameters and flags, data is output to stdout. PyDriller exposes an API, an extension to a programmer's familiar programming paradigm. This is in contrast to ecosystems that define a different abstraction, such as SmartSHARK, where detailed instructions must be followed to extend its functionality \footnote{\url{https://smartshark.github.io/plugin/tutorial/python}}. Extension instructions are also not available for Perceval or Codeface.

\textbf{Licensing.} Another important consideration in reusing a code component is how permissive its license is. For example, stringr, an R package to manipulate strings used by XGBoost, a popular machine learning algorithm, was replaced by stringi, another R package to manipulate strings, solely based on the difference in licenses.\footnote{\url{https://github.com/dmlc/xgboost/issues/1338}}  Similar reasoning also led an R package that represents data tables efficiently to adopt a different license because the existing license ``could be interpreted as preventing closed-source products from using data.table''\footnote{\url{https://github.com/Rdatatable/data.table/pull/2456}}. Lack of clarity on interactions of open source licenses has been reported by \cite{Almeida:2017}. Among the  tools we studied, we have observed the following licenses: Codeface adopts GPL 2.0, PyDriller Apache 2.0, SmartSHARK Apache 2.0, and Grimoire's Lab GPL 3.0 and LGPL 3.0.

\textbf{Perils of Code Reuse.}\label{sec:code_reuse} With the availability of package managers such as CRAN and PyPi which greatly facilitate code reuse, you can declare dependencies on others’ code instead of
copying it into your own project, taking advantage of their functionality without assuming the burden of maintenance. However code interdependence also poses risks \cite{Valiev:2018}, such as dependencies going extinct \cite{Coelho:2017}. Hence, care has to be taken to avoid dependencies to non-maintained third-party  code. 

An interesting example occurs in mecoSHARK\footnote{\url{https://github.com/smartshark/mecoSHARK}} through a chain of dependencies which exemplifies the concern posed here. mecoSHARK is a component that serves as a wrapper for OpenStaticAnalyzer\footnote{\url{https://github.com/sed-inf-u-szeged/OpenStaticAnalyzer}}, with a last commit date of July 13, 2018. In turn, OpenStaticAnalyzer also wraps  several other dependencies, including FindBugs \footnote{\url{http://findbugs.sourceforge.net/}}, last released in March 15, 2015. In its bug tracker\footnote{\url{https://sourceforge.net/p/findbugs/bugs/1487/}}, FindBugs requests for bugs to no longer be reported, noting that SpotBugs\footnote{\url{https://github.com/spotbugs/spotbugs}}, FindBugs' successor, should be used instead. This confirms that the mecoSHARK wrapper, which provides OpenStaticAnalyzer functionality to SmartSHARK,is now dependent on dead code, further increasing the burden of the SmartSHARK ecosystem maintainers. Nonetheless, SmartSHARK's approach to wrap black-box packages into common APIs is considered good practice  \cite{Holt:2015}.

As a means to mitigate this risk, relying on and contributing  work to open source communities that more carefully assess the health of projects and try to maintain them, such as the Apache Software Foundation, ROpenSci\footnote{\url{https://ropensci.org/about/}}, and CHAOSS\footnote{\url{https://chaoss.community/}} may be an important consideration. For example, ROpenSci accepts R packages via a streamlined peer review process and, for accepted packages, provides community support, package promotion, and fast-track publication to journals\footnote{\url{https://devguide.ropensci.org/softwarereviewintro.html\#whysubmit}}.

\section{Design Principles in Kaiaulu}

In this section, we discuss how our design principles are translated into Kaiaulu's specific design decisions. In the following section, we fully flesh out Kaiaulu's modules and features.

\textbf{Batch Mode, Interactive Mode, and Literate Programming in Kaiaulu}. We chose to use the R language\footnote{\url{https://www.r-project.org/}}, due to the familiarity of the authors with the language and a preference for its package architecture. 

Minimally, the structure of an R package consists of the package metadata and its API. In addition, the R ecosystem encourages and promotes best practices to include documentation packages called vignettes, which leads R users to expect an API and R Notebooks when installing packages from CRAN (The Comprehensive R Archive Network).\footnote{\url{https://cran.r-project.org/web/packages/}}  CRAN  treats R Notebooks as first class citizens in an R package\footnote{See for example under Vignettes: \url{https://cran.r-project.org/web/packages/ggplot2/index.html}}  showing on each package's website any R Notebooks available. Because of R package  structure, complying with familiar software abstractions (see Section \ref{sec:software_abstractions})  automatically brings the benefits of literate programming (see Section \ref{sec:batch_interactice_literate}).

\textbf{Abstraction Debt in Kaiaulu.} R natively supports tables and vectors as data types, which is a familiar abstraction for data analysts. To capitalize on this, Kaiaulu's \textit{parse\_} functions map most data sources (Git logs, mailing lists, file dependencies, software vulnerability feeds, metrics,  etc.) as tables with standardized column naming, which allows for quick identification of what data can be combined. Kaiaulu also offers various \textit{transform\_to\_network\_} functions to represent and interactively visualize these networks\footnote{\url{https://github.com/sailuh/kaiaulu/blob/master/R/network.R}} which in turn enable more complex socio-technical analyses at different granularities: functions, files, classes, etc.

\textbf{Tool Configuration Files vs Project Configuration Files in Kaiaulu.} Following the design choice of Codeface (see Section \ref{sec:config_files}), and building on best practices for machine learning configuration files \cite{Holt:2015} we implemented project configuration files using YAML. Because we externalize all parameters in project configuration files, an important concern is that the file does not grow overly complex, requiring documentation of its own. That is, we do not wish the minimal path to data to increase as new features are added, as we discuss next.

\textbf{Minimal Path to Data in Kaiaulu.} As discussed in Section \ref{sec:minimal_paths}, it is important that the path to data remains as simple and short as possible. 
We again build upon familiar concepts, specifically with the intent of applying the rule of least surprise \cite[Ch.11]{Raymond:2003}\footnote{Also publicly available at: \url{http://www.catb.org/~esr/writings/taoup/html/ch11s01.html}} i.e. `do the least surprising thing'. In an R package, it is expected that R Notebooks provide examples of how to leverage the API to accomplish a task by combining multiple functions, while individual functions provide self-contained examples, which can be obtained in the R environment at any time by  preceding a function name with a question mark, e.g. `\textit{?parse\_gitlog}'. 

To build upon this we: 1) \textit{Do not create} any dependency between configuration files and the API: functions take, as input, parameters which are familiar to any programmer; 2) \textit{Use} project configuration files only in the first code block in R Notebooks to load the variables required to use the functions of the API, similar to how best practices in static programming languages encourage variable definitions at the beginning of a program; 3) \textit{Create} a dependency between the CLI and the project configuration files, to facilitate batch processing and reproducibility. 

Our intent is that users will first observe the R Notebooks to get a better understanding of the API for a particular task of interest, and in doing so will familiarize themselves with both the relevant portion of the API and the project configuration file. 
If the interest is only, for example, to understand how to parse Git logs, using for example the Git log R notebook, then users should not be concerned with specifying the mailing list. When comfortable, users can then use their newfound understanding to scale the analysis to the entire project using the configuration file for the CLI, build their own analyses as vignettes, or define new CLI interfaces. This design is consistent with a mining software repositories workflow, in which a researcher should first explore the data qualitatively to assess threats to validity, before scaling up data processing in batch mode without clarity of what assumptions the tool is making using default parameters or arbitrary thresholds.

Kaiaulu also further decreases the minimal path to data in terms of how it handles third party dependencies. Users need only concern themselves with installing dependencies for their task of interest. For example, if the interest is only to parse Git logs, they need only set up Perceval, and provide its binary path as a parameter to Kaiaulu's \textit{parse\_gitlog} to obtain the parsed data. More generally, the \textit{parse\_} API minimizes effort to researchers by transforming various tool-specific data formats, if the researcher so desires, into tables, and performing minimal processing on potentially inconsistent fields, such as file paths, to make them internally consistent.

\section{The Kaiaulu R Package}

Based on the above observations and lessons learned, we now describe the realized modules and features resulting from the design decisions behind the Kaiaulu R package.  

\begin{figure}[!htbp]
 \centering 
 \includegraphics[width=1\columnwidth]{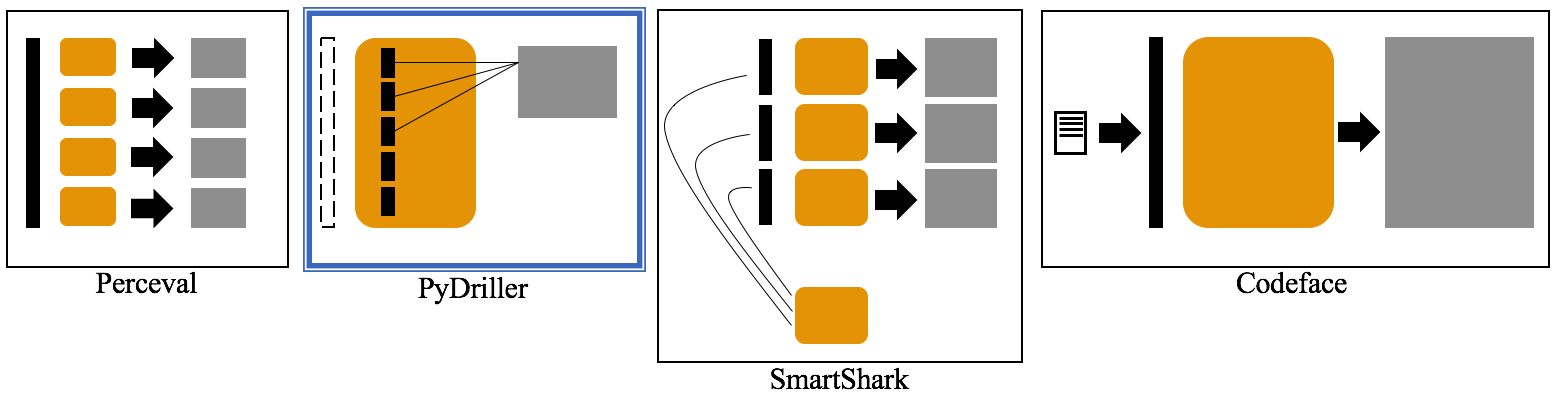}
 \caption{Conceptual diagram of interface, input, and output of tools showcasing differences in design.}
 \label{fig:design_1}
\end{figure}

Mining software repositories often requires the handling of multiple data sources to analyze a project's ecosystem. Minimally, a researcher is required to understand the data source in its native form, acquire it (typically using an API), and parse and save it (e.g. as a table of data). Overheard is incurred if a tool needs to be purpose-built to accomplish these steps. In the best case, the acquisition and parsing steps can be accomplished by using an existing tool. When designing Kaiaulu, we asked ourselves how to emphasize the minimal path to data (as discussed in Sec. \ref{sec:minimal_paths}). To illustrate our rationale, Figure \ref{fig:design_1} revisits some of the tools' design decisions we discussed earlier. 

In Figure \ref{fig:design_1}, Perceval (left) provides a single CLI interface for acquisition of various data sources. For example, a project's issues can be fetched by using the `jira' endpoint, while `git' may be used to parse repositories. Pydriller (center-left), provides functions via a Python API. Users of these tools gain flexibility in parsing the data, at the cost of a higher learning curve and familiarity with the language. SmartShark (center-right) provides similar functionality to Perceval, but endpoints such as `jira' and `git' are now realized as entirely separate tools, orchestrated by another tool. Finally, Codeface (right) provides a single CLI interface, like Perceval. But most of its functionality is executed in batch mode and output into a single database dump, offering the least flexibility in terms of what analyses to execute. Unique to these tools, Codeface stores project parameters in reusable configuration files.

\begin{figure}[!htbp]
 \centering 
 \includegraphics[width=1\columnwidth]{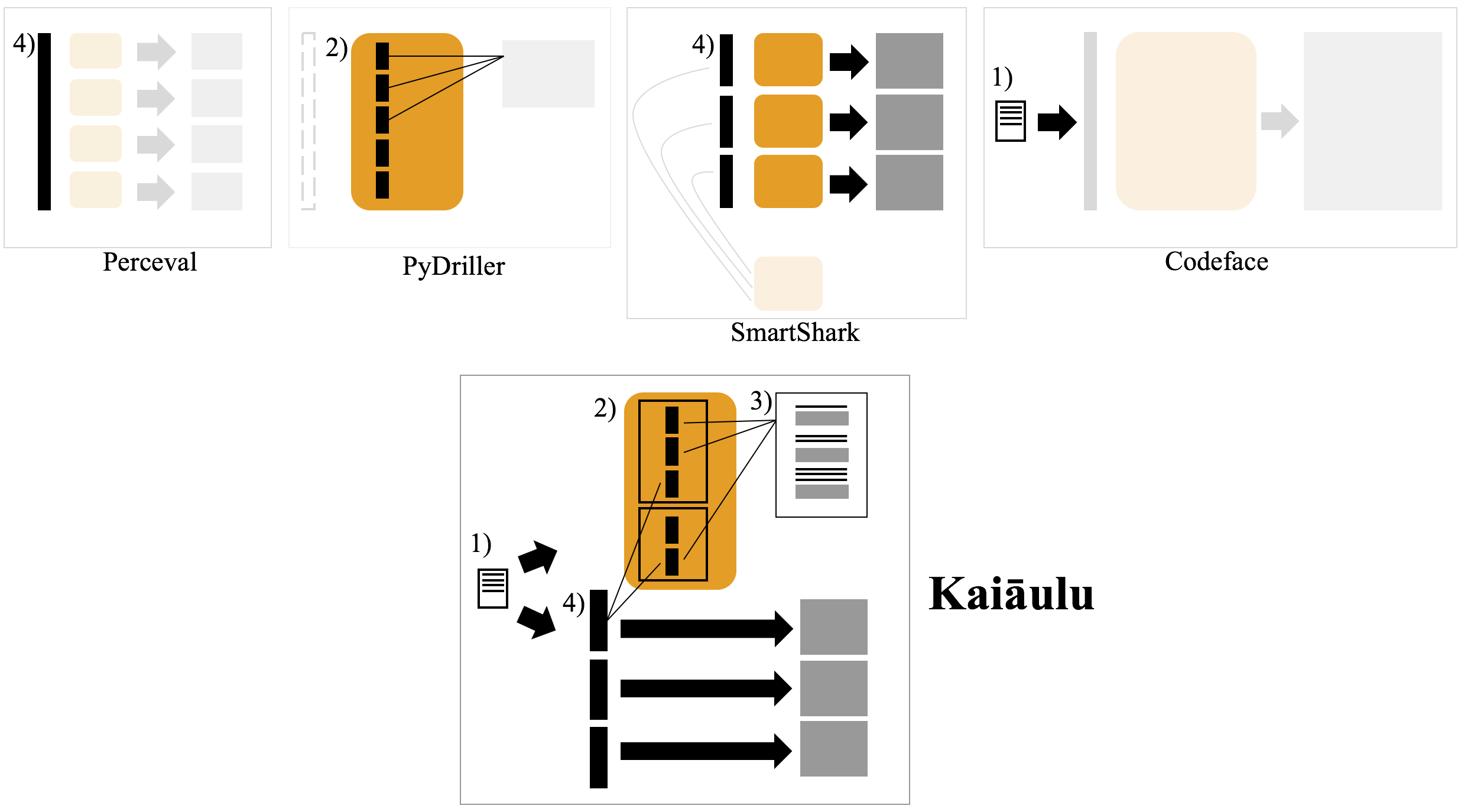}
 \caption{Conceptual diagram of interface, input, and output of tools showcasing how Kaiaulu coompares to the tools.}
 \label{fig:design_2}
\end{figure}

Using Figure \ref{fig:design_1} as a basis for comparison, Kaiaulu's design is shown in Figure \ref{fig:design_2}, separated into parts 1) through 4).  Kaiaulu borrows from the design of PyDriller by defining an API and a set of functions (2). The use of configuration files, inspired by Codeface (3), is done at the R Notebook level (rather than at the function level). That is, project configuration parameters are read into an R Notebook, and appropriate parameters are then passed to functions. This allows us to decouple configurations from function signatures, to tell best practice stories (interspersed with code) of how parameters are used in various analyses \cite{Kalliamvakou:2014,Bird:2009,Bird:2006}, and offer a reusable end-to-end pipeline for specific exploratory analysis. For example, the social smells notebook\footnote{\url{https://github.com/sailuh/kaiaulu/blob/master/vignettes/social_smell_showcase.Rmd}} emphasizes care in assessing project's communication, which are often fragmented over multiple archives. More importantly, R notebooks enable easy manual inspection of intermediate data, such as the use of identity match heuristics. We found this use of `reusable data stories' particularly useful to familiarize undergraduate and graduate research assistants to common pitfalls. 

We borrowed the use of CLIs (4) from Perceval and Smartshark. The CLI serves to accommodate users who are unfamiliar with R; it also supports scaling analyses defined and prototyped in Notebooks, so that they can be run in batch mode. To build upon (2), the CLI simply utilizes the defined API behind the scenes, which simplifies code maintenance. As with the Notebooks (3), the CLI parses  project configuration files, which also facilitate server-side reuse. 

Kaiaulu was designed by combining these concepts from (1-4). It was implemented as an R package, building upon familiar software abstractions. Typically R packages are defined as an API of functions as found in PyDriller and R Notebooks. By showcasing project configuration files where users are expected to learn about the package, users can familiarize themselves with project configuration files and the command line interface, which is less commonly found in R packages, and entirely optional. In the following subsections, Kaiaulu's major processing elements are discussed.

\subsection{Parsers}

In the \textit{Parsers} module, our goal was to minimize a user's effort, in terms of acquiring and parsing project source code. These functions were combined into a single interface with consistent nomenclature.

Each of Kaiaulu's parsers is defined as a function (e.g. \textit{parse\_mbox}, \textit{parse\_gitlog}), which are also accessible via a CLI. We wanted parsers in Kaiaulu to reflect Perceval's philosophy of minimal paths to data, with a small learning curve. That is, given a data source, we would like users to quickly be able to see the data without spending excessive time on setup. As such, each parser function is given a single responsibility: to display a data source as a table with a standardized column nomenclature (in case multiple sources referred to the same data with different names). Unlike Perceval, since an API option is also available, users can interactively prototype and analyze the data in the R environment. Having tables as the default output option minimizes the time spent learning what fields are available in the source. The standardized nomenclature allows for intuitive joining operations across the outputs of different parsers. 

To account for \textit{the perils of code reuse}, Kaiaulu limits its interface only to third party software that have CLI interfaces. Parsers with third party dependencies simply contain in their signatures an additional parameter for path to the required binary. This dependency mechanism allows users to bypass setting up third party tools which they do not directly need to use. Additionally, users benefit from using the Kaiaulu function to obtain a tabulated and standardized data input. For example, the \textit{parse\_gitlog(git\_repo\_path,perceval\_path)} function requires, as input, Perceval's binary to tabulate its JSON output. In this way parsers can build upon third party functionality to implement new features. For example, \textit{parse\_gitlog\_entity(git\_repo\_path,utags\_path,project\_git\_log,kinds)} implements a git log parser capable of tabulating 
developer changes from git logs at the granularity of functions rather than files (inspired by Joblin et al \cite{Joblin:2015}). In addition, the assumptions and threats to validity in the cited work are provided in the Notebook \footnote{\url{https://github.com/sailuh/kaiaulu/blob/master/vignettes/blamed_line_types_showcase.Rmd}}. 

While simple in concept, we note that existing tools do not offer this functionality. For instance,  Codeface \cite{Joblin:2015}  offers an implementation of a function-based git log parser, but since it has an `all-in-all-out' interface, this function can not be reused elsewhere. The same is true for SmartShark. Perceval, while containing a shorter path to data, still requires tabulation and standardization of the collected results. Lastly, PyDriller does not adopt the philosophy used here for extending functionality based on third party software, as it limits its scope to Git. 

Kaiaulu currently employs a variety of parsers, providing the ability to parse git logs, mailing list archives (e.g. pipermail, Apache's mod\_mbox), issue trackers (e.g. Jira, GitHub), static parsers (file and function dependencies), evolutionary parsers (file and function changes), commit hashes (e.g. to identify issue ids from commit messages), and software vulnerability feeds. Parsers which contain filepaths also contain optional regular expression filters to whitelist or blacklist files based on their extensions or naming conventions. For example, we use this to remove test files from analyses as these files could compromise code metrics.

\subsection{Transformers, Graphs and Networks}

Kaiaulu's Transformer, Graph, and Network modules are grounded on the observation that most software and social metrics are graph-based (e.g. co-change, fan-in, fan-out, communication). These modules \textit{transform} the data provided by various kinds of parsers that parse the raw project data. Transformers reformat the data provided by parsers into lists of nodes and edges which are then represented as networks, using \textit{graph} data structures. In this way we can more easily visualize and explore the \textit{networks} of relationships among a software project's elements.   

Kaiaulu represents the socio-technical network for each snapshot as a graph $G_{st} = (V, E)$, where the set of nodes $V = V_a \cup V_f \cup V_t$ comprises authors $V_a$, source files $V_f$ and e-mail threads $V_t$. The set of edges E = $E_{comm} \cup E_{chg}$ models communication and collaboration between authors, where communication is done via $E_{comm} \subseteq V_a \times V_t$, and file changes via $E_{chg} \subseteq V_a \times V_f$. Observe by this construction, the socio-technical network is in fact two bi-modal bipartite networks $G_{st} = G_{chg} \cup G_{comm}$. Both $G_{chg}$ and $G_{comm}$ are also weighted (representing an author's count of changes to a file within a user specified time window (e.g. 3 months), and the number of replies submitted to an e-mail thread respectively), and undirected (the direction is irrelevant in this case because it could only go in one direction in each bipartite network). Likewise, the CVE and File Networks are weighted, undirected, bipartite graphs. The definition of various transformations are encapsulated separately in functions, consistent to the overall architecture. 

\textbf{Projection Transformations.} 
Familiar software engineering metrics can be derived from graph projections. Intuitively, a graph projection operation eliminates one set of the `colored' nodes in a bipartite graph, and connects the adjacent black nodes together, where the resulting edge weight is the sum of the eliminated edges. For instance, in a bipartite network represented by file and commit nodes, eliminating the commit nodes would result in a file's co-change metric, revealing \textit{indirect collaboration}. For example, if five authors modified the same file within a given time period, then the projection operation shows that \textit{all five authors indirectly collaborated}, irrespective of the order of their changes (note that the derived uni-modal networks are undirected). We define this as the projection transformation to go from bi-modal to uni-modal networks. 

\textbf{Temporal Transformations.} Let us now consider a second approach to obtain uni-modal networks. In \cite{Joblin:2015}, the authors define one method to construct uni-modal networks from the same data by defining indirect collaboration using the notion of incremental contributions through the timestamps on commits. For example, if author A modifies a file, and very next change to the same file is performed by author B, then B is said to have \textit{indirectly collaborated with A}. A similar intuition and transformation could be used to categorize e-mail replies. We define this method, to go from the bipartite network to the uni-modal network, a \textit{temporal} transformation (as it relies on the timestamps). Observe in this case that the derived uni-modal networks will be directed graphs (which indicate the flow of time). The edge's weight is defined as the sum of lines of code added by both developers in their respective file changes.

\begin{figure}[thb]
    \centering
	\includegraphics[ clip,width=1\linewidth]{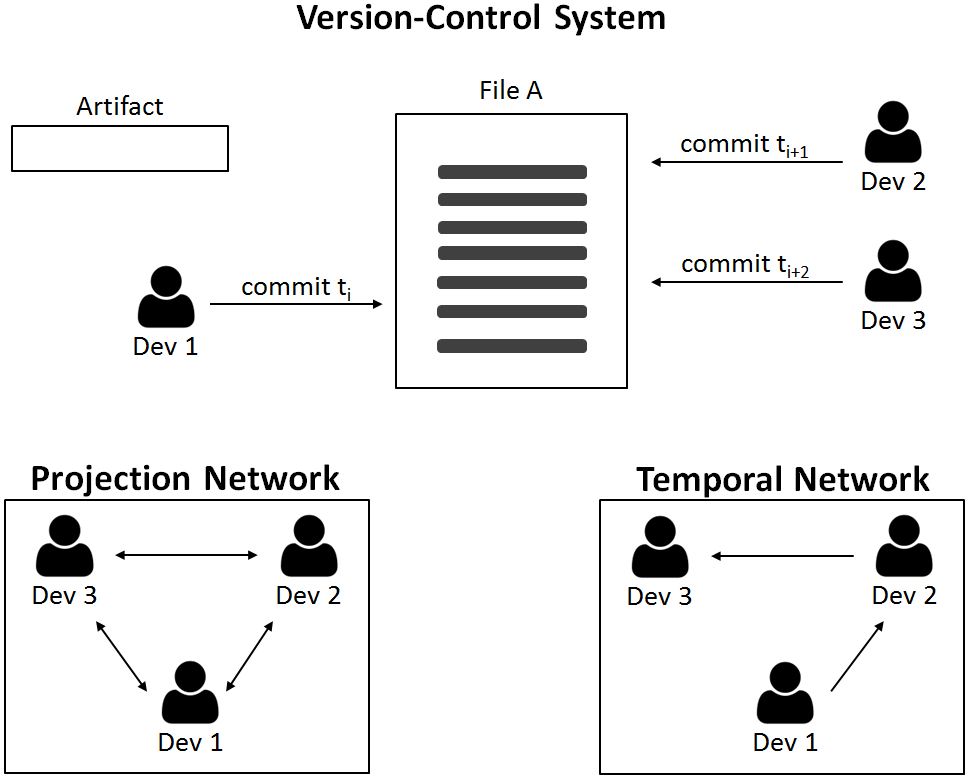}
\caption{Temporal vs Projection Networks, format adapted from \cite{Joblinthesis:2017}.}.
\label{ch6_fig:temporal_vs_projection}
\end{figure}

Which method to derive uni-modal networks should we choose? This decision is encapsulated in Kaiaulu by the choice of functions. By swapping projection and temporal transformation functions, users can experiment with, and visualize, their various implications.  

\begin{figure}[thb]
    \centering
	\includegraphics[ clip,width=1\linewidth]{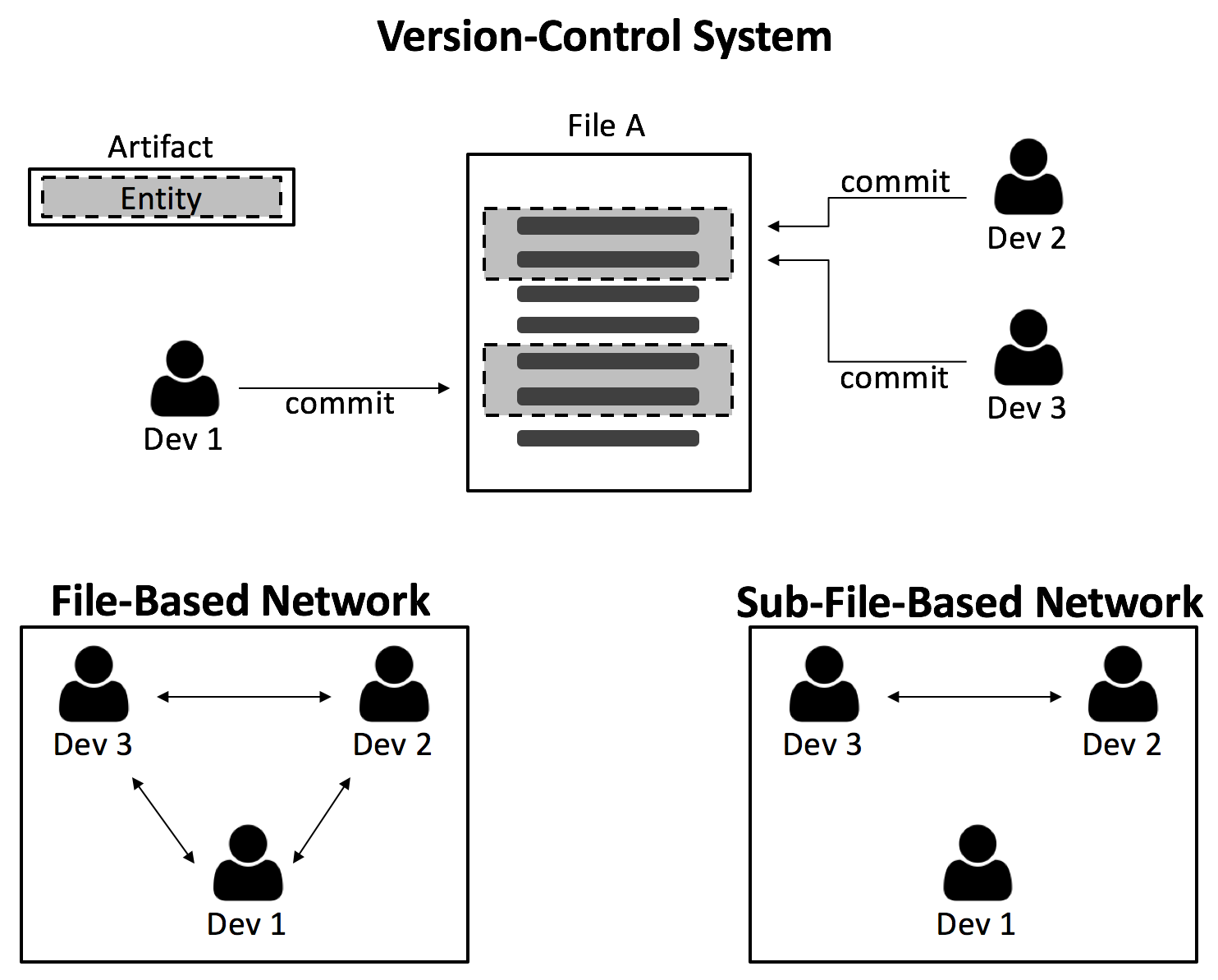}
\caption{File vs Sub-file (e.g. Function) Networks, adapted from \cite{Joblinthesis:2017}.}.
\label{ch6_fig:file_vs_function}
\end{figure}

\begin{figure*}[ht]
  \subfloat[CVE and File Network.]{
	\begin{minipage}[c][1\width]{
	   0.5\textwidth}
	   \centering
	   \includegraphics[width=1\textwidth]{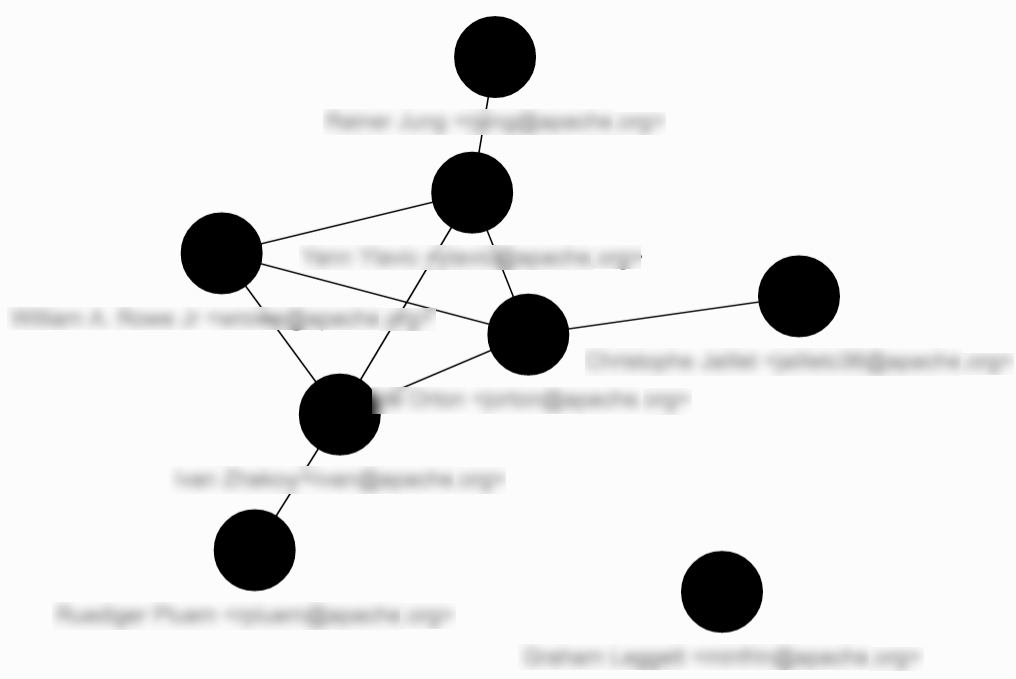}
	\end{minipage}}
 \hfill 	
\subfloat[Projection Network. Nodes indicate developers. Edges represent common changed files.]{
	\begin{minipage}[c][1\width]{
	   0.5\textwidth}
	   \centering
	   \includegraphics[width=1\textwidth]{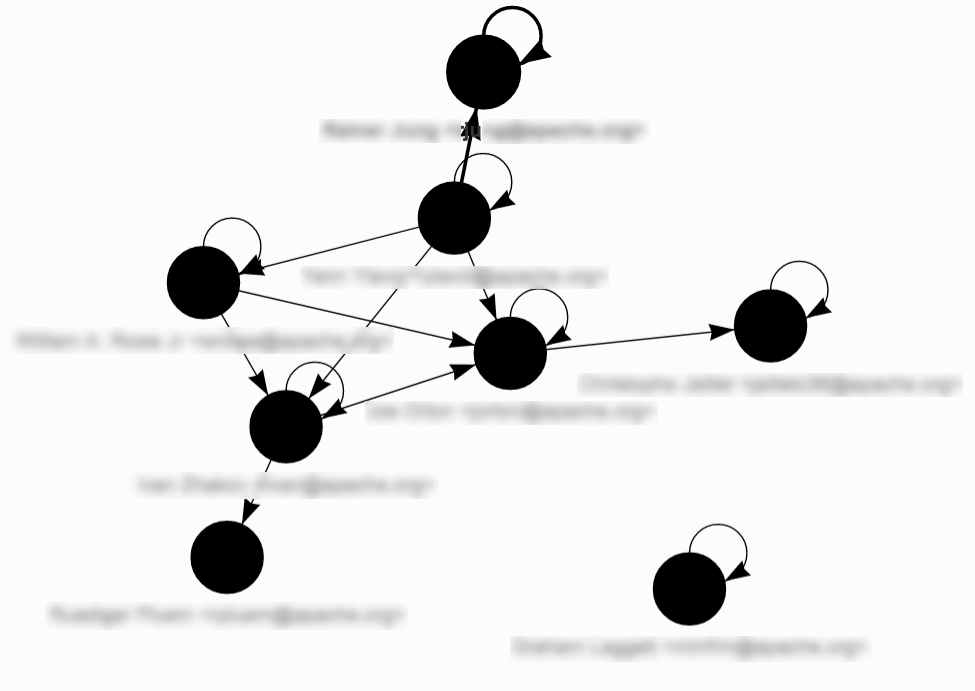}
	\end{minipage}}
\caption{Temporal Network. Nodes indicate developers. Edges represent common changed files. The edge direction indicates the temporal order of change.}
\label{projection_temporal_example}
\end{figure*}

An example of both projection and temporal networks is shown in figure \ref{projection_temporal_example} (the name of each node's developer has been blurred). In the projection network, developers are connected if in a given time window they modified any file in common. In the temporal network, the direction displays which developers changed files after which in the given time window. For example, a bidirectional arrow means two developers change the same file together. A uni-directional arrow suggests another developer may have ``taken over'' during that time window. In both cases, we could derive hypothesis of the nature of collaboration, and derive hypothesis to be tested in the exploratory analysis. 

\textbf{File, Functions, and Entities.} Another consideration  encapsulated in Kaiaulu's functions is which entities are analyzed to derive indirect collaboration. For example, consider Figure \ref{ch6_fig:file_vs_function}. As we can see, the choice of granularity will also affect the number of edges generated, where a file granularity generates more edges than function granularity. A larger number of edges, in turn, may impact the social smell metrics, as the existence of connections between developers in one network, and the absence of edges in another network, may inflate the count of metrics, such as social smells (which we define in the next section). The authors in \cite{Joblin:2015} used a combination of temporal transformation and function granularity. In Kaiaulu, we implemented both the file granularity, and generalized the function granularity to \textit{entities}, where an sub-file unit can be any source code block region of interest (e.g. functions, classes, or language specific features like structs in C). 

\subsection{Identity} 

A critical component of conducting socio-technical analysis in open source communities is assigning a consistent identity to users who may employ multiple variants of their name and e-mail addresses in their project interactions. Several approaches to match identities have been proposed (e.g. \cite{Bird:2006}, \cite{Zhu:2019}). Exact name matching (either names or e-mails) or partial matching (e.g. based on edit distance) are two commonly used schemes. 

Our identity matching was designed as a 3 step pipeline: formatting, name-email separation and pair-wise matching. Formatting includes the removal of symbols such as `$<$ $>$', commas or replacing `at' with `@', while avoiding modifying a name, such as Matt. Name and e-mail separation handles cases where first or last or both names are not provided, multiple word names, etc. Finally, pair-wise matching handles comparisons of name and e-mail, or reversed names. 
In total, the steps of formatting, name separation, and name matching amounted to 31 test cases, which were successfully implemented. At the end of this step, users in  the version control system, issue tracker, and mailing list who matched via the tests we implemented were assigned an appropriate ID. Thus, given the name, and optionally the e-mail, other information sources can be matched.

\begin{figure}[!htbp]
	   \includegraphics[width=1\textwidth]{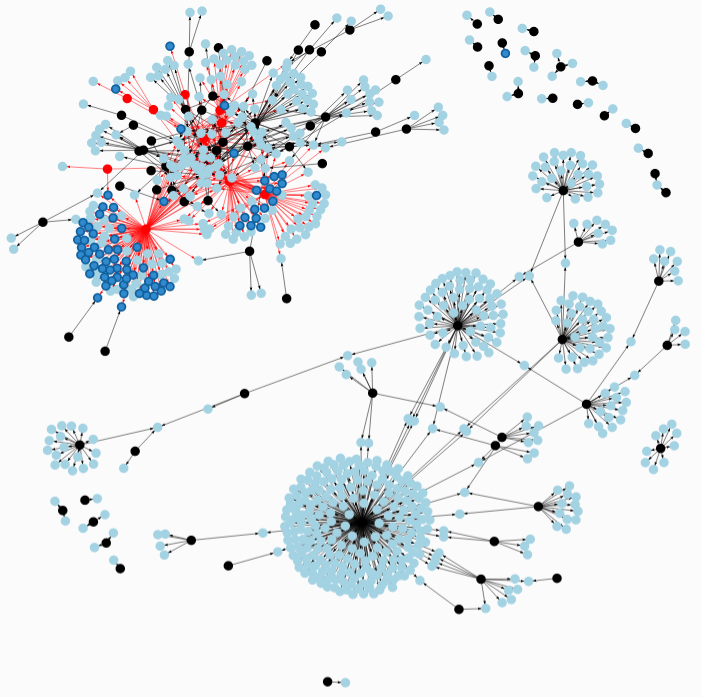}
\caption{Reply networks combine communication networks. Here dark blue nodes are issue comments, light blue nodes are mailing list comments, and black nodes are developers. Red nodes are developers who communicate in both the mailing list and the issue tracker.}
\label{reply_network}
\end{figure}	

An example of the utility of identity matching is shown in figure \ref{reply_network}. Here, project communication occurs in parallel in both the Jira issue tracker and the project's mailing list. We fuse these information sources into a single ``Reply Network".

In summary, the transformer API  provides users with flexibility with respect to both temporal assumptions and sub-file granularity. Because all networks are  annotated graphs, community detection algorithms in Kaiaulu can be used to identify important patterns. For example, if applied to a file-commit network over a fixed period of time, co-changed file clusters can be identified. Similarly, if the file network is derived from file to file dependencies, clusters related to modularity measures can be derived. Developer networks can be used to detect communities. 

\begin{figure}[thb]
    \centering
	\includegraphics[ clip,width=0.8\linewidth]{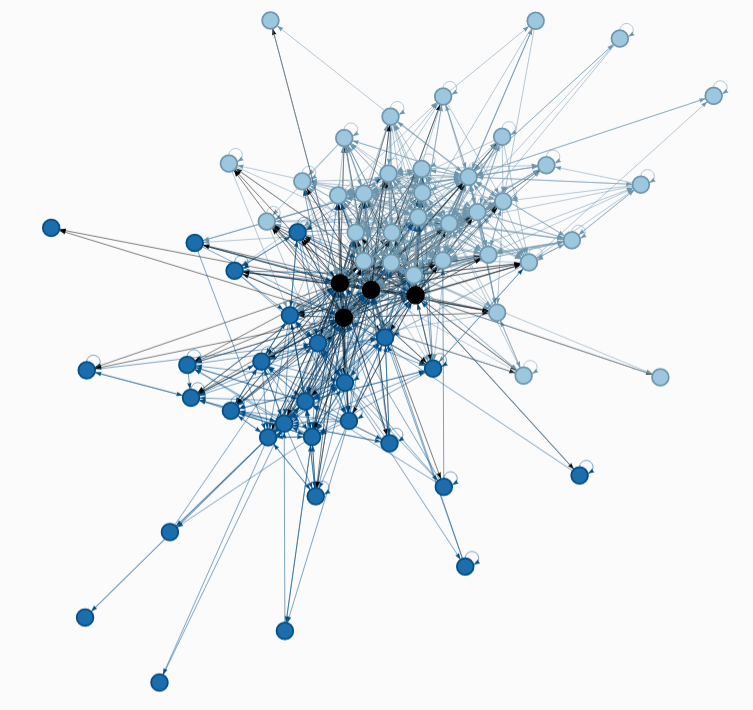}
\caption{Community detection applied to temporal projection.}.
\label{community_detection}
\end{figure}

In figure \ref{community_detection}, we apply community detection to a temporal network such as the one illustrated in figure \ref{projection_temporal_example}. The result is displayed by re-coloring the black nodes. Darker blue and lighter blue nodes represent two communities of developers as determined by the files that they changed in common. Developers in black represent boundary nodes, which participate of both communities. In the interactive format, researchers can ``zoom in'' on these nodes to identify who are the common developers, and can use this information to draw further hypotheses.

\subsection{Metrics}

In the metrics module, we define some commonly used  metrics, such as number of bugs, churn, LOC, as well as the less well-known social metrics. Demographics are also provided to help contextualize the previously presented social networks, such as the number of developers modifying files and exchanging e-mails, number of files, threads, and different timezones\footnote{For a full analysis with the metrics, see: \url{http://itm0.shidler.hawaii.edu/kaiaulu/articles/social_smell_showcase.html}. The source code can be found on \url{https://github.com/sailuh/kaiaulu/blob/master/vignettes/social_smell_showcase.Rmd}}.

\begin{figure}[thb]
    \centering
	\includegraphics[ clip,width=0.8\linewidth]{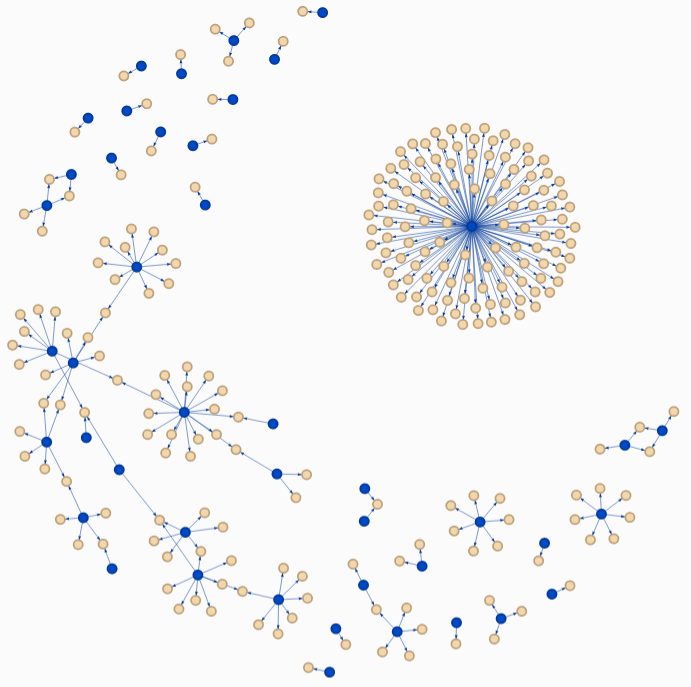}
\caption{Issue Network. Blue nodes represent issues, and yellow nodes files}.
\label{issue_network}
\end{figure}

For bug counts, rather than simply interpreting these as metrics, we make it easy for users to observe their topology. This is evident in figure \ref{issue_network}, where a single issue is associated with multiple files (top right). While some other files may have a lower count of issues, more complex structure (such as we can observe at the bottom left of the figure) would be missed if only metrics were employed. Furthermore, feature issues can be discerned from bugs by examining the issue labels.

We devote this section to briefly discuss the social metrics as they leverage the previously discussed modules. For the social metrics, we adopt the definitions of social smells defined by \cite{Tamburri:2019}.
Social smells reflect recurring sub-optimal organizational structure patterns connected to organizational behavior patterns, e.g., sub-optimal knowledge sharing, recurrent sharing delays, misguided collaboration and more.  We chose to integrate three of these smells---Organizational Silo, Missing Link and Radio Silence---and two related metrics: socio-technical congruence and missing communicability \cite{Tamburri:2019}. 
Here we explain one of the social smells; additional details about these metrics can be found in \cite{Tamburri:2019}. 

\begin{figure*}[htbp]
\centerline{\includegraphics[scale=0.39]{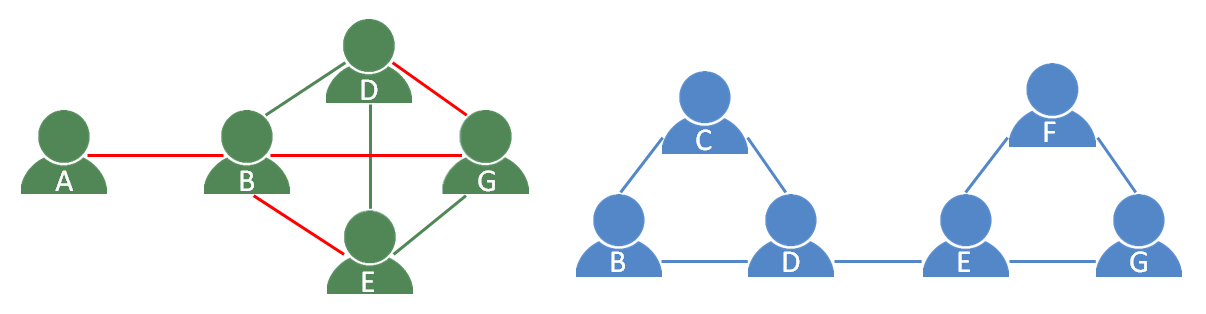}}
\caption{Missing Link Social Smell \cite{Simonethesis:2016}}
\label{ch6_fig:missing_link_social_smell}
\end{figure*}

In Figure \ref{ch6_fig:missing_link_social_smell}, the collaboration network projection is shown in green to the left. The communication network is shown in blue to the right. The intent behind developer networks is to capture developers who modified the same file in a given snapshot, and also communicated via a common e-mail thread in a user-specified time window (e.g. 3 months). In this example, we can see to the left highlighted in red that developers (A,B), (B,E), (B,G), and (D,G) collaborated (i.e. they have a red edge in the green graph), but do not communicate (they do not have an edge in the blue graph). Therefore, the missing link social smell is counted 4 times for this snapshot.

\subsection{Configuration}

Currently, the configuration module is minimal and exists embedded in R Notebooks. As noted at the start of this section, we borrowed  from Codeface the idea of project configuration files. And we include more parameters in the configuration file as compared to Codeface. For example, Codeface hardcodes the set of acceptable file extensions, whereas Kaiaulu defers this choice to the user in the project configuration file. 

More generally, we adopt the philosophy that project configuration files should serve as a distilled representation of assumptions and analysis choices. Rather than just serving as a repository of configuration choices for reproducibility, it is readable as plain text, and so can be easily exchanged and discussed. In Kaiaulu, a project configuration file is written in YAML. An example of project configuration file can be found in the public tool repository `conf` folder\footnote{\url{https://github.com/sailuh/kaiaulu/tree/master/conf}}.

As with every design decision in Kaiaulu, the full project configuration file needs not be specified. Indeed, every R Notebook, at the beginning, clarifies which parameters are required. In future work, we plan to expand the Configuration module to tabulate multiple configuration files. For example, it is often common in software engineering literature to analyze multiple projects and present a summary statistics table of the projects to assess generalization of results. Such tables could be generated on the fly from the files. Ideally, project configuration files should suffice as supplementary material, alongside Kaiaulu's version for full reproducibility.

\section{Conclusions and Future Work}
In this paper, through an action research approach, we have determined a set of key design decisions mined from existing tools. Based on these lessons learned we iteratively developed Kaiaulu, an R package for mining software repositories.  Our goal in creating Kaiaulu was to simplify most of the boring, repetitive, and error-prone tasks in mining software repositories, leaving the user free to focus on the true goals of their research.

In Kaiaulu we have implemented and released a comprehensive set of capabilities to mine, analyze, and visualize software repositories, including social smells \cite{Tamburri:2019}, architecture smells and metrics \cite{Mo:2019}, and bug timelines  based on prior work by other authors. Kaiaulu is licensed under MPL 2.0.

While we have derived the principles for Kaiaulu from our action research, our future work is to take a more disciplined approach to Kaiaulu's design, based on the quality attributes that represent its architectural drivers.  Following the approach outlined in \cite{Kazman:2021} and \cite{Bass:2021} we can, in the future, attempt to more systematically collect architectural drivers, make reasoned design decisions, and support these decisions with well-established design rationale. 

%
%
%
 \bibliographystyle{splncs04}
 \bibliography{main}

\end{document}